
\def\today{\ifcase\month\or
  January\or February\or March\or April\or May\or June\or
  July\or August\or September\or October\or November\or December\fi

  \space\number\day, \number\year}

\magnification \magstep1

\footline={\hfil}
\rightline {DAMTP R93/33}
\rightline { hep-th/9312014 }
\vskip 1cm

\centerline{ \bf Flux-confinement in Dilatonic Cosmic Strings }
\vskip 1cm
\centerline { G W Gibbons\footnote{$^1$}{e-mail address GWG1@AMTP.CAM.AC.UK}
and C G Wells\footnote{$^2$}{e-mail address CGW11@AMTP.CAM.AC.UK}}
\vskip 1cm
\centerline { DAMTP}
\vskip 1cm
\centerline {University of Cambridge}
\vskip 1cm
\centerline {Silver Street}
\vskip 1cm
\centerline {Cambridge}
\vskip 1cm
\centerline {CB3 9EW}
\vskip 2cm
\input mssymb

{
\narrower
{\bf Abstract}
We study dilaton-electrodynamics in flat spacetime and  exhibit a set of global
cosmic string like solutions in which the magnetic flux is confined. These
solutions continue to exist for a small enough dilaton mass but cease to do
so above a critical value depending on the magnetic flux. There
also exist domain wall and Dirac monopole solutions.
We discuss a mechanism whereby magnetic monopoles
might have been confined by dilaton cosmic strings during an epoch in the
early universe during  which the dilaton was massless.
}

\vskip 1.5cm
\centerline{\it Submitted to Classical and Quantum Gravity}

\vfill
\eject
\count0=1
\footline={\hss\tenrm\folio\hss}
\headline{\centerline{Flux-confinement in Dilatonic Cosmic Strings}}

\beginsection 1. Introduction

There has been considerable interest recently in 4-dimensional
Dilaton-Einstein-Maxwell and Dilaton-Einstein-Yang-Mills theory. In the
abelian case single and multi
static and stationary black hole
solutions with electric and magnetic charges have been extensively studied.
Another class of interesting solutions describe magnetic
fields with or without black holes. For Einstein-Maxwell theory
these are based on the Melvin solution which represents a sort of
super-massive cosmic string [1,2]. The Melvin
solution may be generalized to include a coupling to the dilaton [3].
Related black hole metrics have been
obtained and their relevance as instantons for the Schwinger
production of black holes discussed [4]. Recently Maki and Shiraishi [5]
have obtained some interesting time-dependent solutions with a dilaton
potential.

In this paper, in
order to gain some physical insight into dilaton-electrodynamics and its
non-abelian generalization,  we will study the simpler
flat-space version in which the effects of gravity are ignored. The action is

$$
\int  d ^4 x \Bigl( - {1 \over 4} \exp (-2 \tilde \kappa
 \phi )
F_{\mu \nu} F ^{\mu \nu} - {1 \over 2} \eta ^{\mu \nu} \partial _{\mu} \phi
 \partial _{\nu}
\phi   \Bigr)
\eqno (1.1)
$$
where $\eta ^{ \mu \nu} = {\rm diag} (-1,+1,+1,+1) $ and the dimensionful
quantity $\tilde \kappa$ in the action can be changed by
a suitable rescaling
of the variables and so its numerical value (as long as it does not vanish)
has no physical significance in the purely classical theory which we study
here.
The action (1.1) can be obtained from the standard gravitational case, in
Einstein conformal gauge,
in which the Maxwell field is multiplied by $\exp (-g \sqrt { 4 \pi G} \phi )$
by letting Newton's constant $G$ tend to zero while at the same time
$g \sqrt {4 \pi G}$ tends to the finite limit $\tilde \kappa$ \footnote {*}
{The sign of $g$ we are using in this paper is consistent with the conventions
of the majority of string theorists though it differs from that used in our
recent preprint [6].}. The field
$F_{\mu \nu}$ may  be abelian or non-abelian. In the latter case there remains,
again in the classical theory, sufficient freedom to scale the Yang-Mills
connection $A$  such that $F=dA + A \wedge A $ . Note that had we taken the
above limit of the gravitational Lagrangian expressed in
string conformal gauge
(i.e in terms of the metric $\exp(2 \tilde \kappa \phi) g_{\mu \nu}$)
we would have obtained a different action from (1.1).

The flat space version (1.1) has been studied by  Lavrelashvili and Maison [7]
and also by Bizon [8] who have obtained sphaleron
type solutions in which the Yang-Mills field is confined by
the attractive forces exerted by the dilaton which replaces the attractive
forces
due to gravity in the Bartnik-McKinnon solution [9].
In the abelian case it is easy
to obtain the general static spherically symmetric
purely magnetic solution representing a Dirac monopole
coupled to the dilaton. Bizon has  pointed out
a special case of the spherically symmetric Dirac monopoles
may be generalized to give multi-Dirac-monopole
solutions. These are Bogomol'nyi type solutions
and may be regarded as a limiting case of the
multi black hole solutions as we shall show in section 4. It is also
completely straightforward to obtain plane wave solutions in which
the dilaton and the photon are travelling parallel to one another.

The main point of this paper is to derive cosmic string like solutions.

\beginsection 2. Permeabilities and Permittivities

It follows  immediately from (1.1) that the equations of motion for
the field $F$ in the presence of the dilaton field are those for a medium in
which the electric
permittivity $\epsilon$ is given by
$$
\epsilon = \exp(- 2 \tilde \kappa \phi)  \eqno(2.1)
$$
and the magnetic permeability $\mu$ is given by
$$
\mu = \exp(2 \tilde \kappa \phi). \eqno(2.2)
$$
The product $\epsilon \mu$ is unity and so the velocity
of light remains one everywhere.
Thus regions of spacetime for which $\phi <0$ are diamagnetic  while
regions with $\phi > 0$ are paramagnetic. One does not usually encounter
permittivities $\epsilon$ which are less than unity. One exception
is the case of vacuum polarization in quantum chromodynamics
which is \lq \lq anti-screening\rq \rq rather than \lq \lq screening\rq \rq
as for all conventional material media. In the non-abelian theory
$\exp (\tilde \kappa \phi)$ plays the role of spacetime dependent  gauge
coupling \lq\lq constant\rq\rq
and in string perturbation theory its expectation
value
plays the role of a variable coupling constant. Thus weak coupling corresponds
to  a diamagnetic phase and strong coupling to a paramagnetic phase. The
action (1.1) is invariant under the simultaneous change  of
the sign of the dilaton field $\phi$ and the replacement of the Maxwell field
$F$ by its Hodge dual.
This symmetry therefore interchanges the weak and the strong coupling phases.

The equation for $\phi$ in the static case is
$$
\nabla ^2 \phi =  -\tilde \kappa  \exp (-2 \tilde \kappa \phi)
\bigl ( {\bf B}^2 - {\bf E}^2 \bigr ), \eqno(2.3)
$$
where $\bf E$ and $\bf B$ have their usual meaning. It follows from (2.3)
that $\phi$ can have no maximum in a purely magnetic field
and no minimum in a purely electric field. Thus if $\phi$ is
taken to be zero at infinity then magnetic regions  tend to be
paramagnetically polarized ($ \mu > 1 $) and electric regions tend to be
dielectrically polarized ($\mu < 1$).  Intuitively magnetic
flux ($\int {\bf B}.{\bf dS}$)
 tends to get self-trapped in strong
coupling domains and electric flux ( $\int {\bf D}.{\bf dS}$) in weak coupling
domains, where ${\bf D}= \epsilon {\bf E}$ is the electric displacement
and (for later use) ${\bf H} = \mu^{-1} {\bf B}$ is the magnetic induction.
These observations
are borne out by the particular solutions mentioned above. Thus for
the Maison-Lavrelashvili-Bizon sphalerons  $\mu$  has a maximum at the centre
and decreases monotonically to unity at infinity. For
magnetic black hole solutions $\mu$ increases monotonically inwards from unity
at infinity. In the extreme case it becomes infinitely large and positive as
one approaches the horizon. From the string point of view the infinitely long
throat is a strong coupling region. In the electrically charged case the
opposite is true. There is a parallel here with monopoles in Yang-Mills theory
and vortices in the abelian Higgs theory. In the case of
$SU(2)$ Yang-Mills theory with the Higgs in the adjoint representation one may
take the  components of the Higgs field as a triplet of permittivities.
The parallel is not completely precise but it is the case for the
't~Hooft-Polyakov monopole that the associated permeabilities monotonically
increase as one moves radially inwards. Similarly for the Nileson-Olesen
vortex one may think of the magnetic flux as being confined inside a core of
high permeability where the Higgs field has a smaller magnitude than it does
at infinity.
We shall see similar features arising for dilaton electrodynamics.

The paramagnetic behaviour described above and the existence of
the dilaton-Melvin solution strongly suggest that there may be
non-singular static cosmic string type solutions in which
a finite amount of flux is trapped. This is indeed the case, as we shall show
in the next section.

\beginsection 3. Dilaton Cosmic Strings and Domain Walls

In the static case the dilatonic Maxwell equations are readily seen to be
satisfied if the magnetic induction
${\bf H}= (0,0, H)$, where $H$ is a constant so long as the dilaton satisfies
the two-dimensional Liouville equation:
$$
\nabla ^2 \phi = -\tilde \kappa  H^2 \exp (2 \tilde \kappa \phi)  \ .
\eqno(3.1)
$$
The general solution of (3.1)
is
$$
\mu = \exp (2 \tilde \kappa \phi) = { 4 \over {\tilde \kappa ^2 H^2 }}
{ { f^ {\prime}(\zeta) g^ {\prime} ({\overline \zeta } )} \over
{\left (1 +    f(\zeta) g( {\overline \zeta}) \right )^2 }} \ \ ,\eqno(3.2)
$$
where $f$  and $g$ are locally holomorphic functions and $\zeta = x+iy$. If
$\phi$ is real then
$$
 {\overline {g  \left ( \> { \overline \zeta} \> \right )}} /f(\zeta)
$$
must
be a real valued holomorphic function of $\zeta$ and hence constant. With no
loss of generality this constant may be taken to be unity and therefore
 the solution we require is
$$
\mu = \exp (2 \tilde \kappa \phi) = {4 \over { \tilde \kappa ^2 H^2}}
{{\left | f^ {\prime}(\zeta) \right | ^2 } \over
{\left (1 +  \left | f(\zeta) \right |^2 \right )^2 }} \ \ . \eqno(3.3)
$$
Note that $f$ and ${ 1 / f}$ give the same solution $\phi$.

Choosing different functions $f$ gives different types of solution. For example
choosing $f(\zeta )=\zeta $ gives a cylindrically symmetric solution with
finite total magnetic flux:
$$
\Phi = \int _{{\Bbb R} ^2} {\bf B}.{\bf dS} = {{4 \pi} \over {\tilde \kappa ^2
H}} \ \ . \eqno(3.4)
$$
This solution may be obtained as limit of the dilaton-Melvin solution.

The magnetic contribution to the total energy per unit length
of our solution is ${1 \over 2} \Phi H =2 \pi /{\tilde \kappa ^2}$ which
is independent of the magnetic field $H$. The dilaton however contributes
a logarithmically divergent energy per unit length because of its
logarithmic dependence on the radius. In this respect our solution resembles a
global
rather than a local string. However it should be pointed out that the
solution, in common with its gravitational version,    breaks neither
electromagnetic gauge invariance
nor any compact internal symmetry.

In addition to the single string solution there are multi string solutions.
Thus choosing a rational function  of Brouwer degree $k$, i.e. the ratio
of two polynomials of order $p$ and $q$,  gives a solution with finite total
magnetic flux
$$   \Phi = k {{4 \pi} \over {\tilde \kappa ^2 H}} \ \ . \eqno(3.5) $$

Note that the permeability decreases to zero as $(x^2 +y^2 )^{-(|p-q|+1 )}$ at
infinity. Thus  the weak coupling region at infinity
is strongly diamagnetic
and  confines the magnetic flux $\Phi$. The multi-string solutions are
not axisymmetric. In flat-space Maxwell theory the only regular solution
is the uniform magnetic field which is necessarily axisymmetric. When gravity
is included this
goes over into the Melvin solution which is also
has axial symmetry. If one insists that
the metric be boost-invariant then the axisymmetry and hence uniqueness,
follow by a version of
Birkhoff's theorem [10]. However the proof given in [10] does not go through in
the presence of a scalar field. This suggests that there may exist static
non-axisymmetric multi-dilaton-Melvin solutions.


If we take $f(\zeta) = \exp(\zeta)$ then
$$
\mu = \exp(2 \tilde \kappa \phi ) = { 1 \over {\tilde \kappa ^2 H^2}} { 1 \over
{\cosh ^2 x }} \ \ .\eqno(3.6)
$$
This solution describes a sheet or membrane confining an amount of flux per
unit length $2 \over {\tilde \kappa ^2 H}$. It may be thought of as a sort of
domain wall separating two
weak coupling domains.

\beginsection 4. Monopoles and Bogomol'nyi Solutions

In this section we shall consider a general static magnetic solution.
Since
$$
\nabla \times  {\bf H} =0 \ , \eqno(4.1)
$$
we may locally introduce a magnetic potential $\chi$ by
$$
{\bf H} = \nabla \chi \ .  \eqno(4.2)
$$
If we make the {\it ansatz}
$$
{1 \over {\tilde \kappa}} \exp ( - \tilde \kappa \phi) =  \chi  \eqno(4.3)
$$
then all the equations will be satisfied if in addition
$$
\nabla ^2  \exp ( \tilde \kappa \phi ) =0 \ . \eqno (4.4)
$$
These solutions are the same as those mentioned by Bizon and may be obtained
from the multi black hole solutions by a limiting procedure.
Our cosmic string solutions are not a special case.

In addition to these Bogomol'nyi solutions it is easy to find
the general spherically symmetric monopole solution. If one insists that
$ \phi$ does not blow up at finite non-zero radius or at infinity
one finds that
$$
\exp (\tilde \kappa \phi)= { 1 \over \alpha } \sinh \alpha \tilde \kappa P
\left ( { 1\over r} + {1 \over b} \right ) \eqno (4.5)
$$
and
$$
{\bf B} = {P \over {r^3}} {\bf r}, \eqno(4.6)
$$
where the constant of integration $b$ is chosen so that $\phi = 0$ at infinity
and $P$ is the total magnetic charge. Just as in the case of magnetic black
holes and 't~Hooft-Polyakov monopoles we find that the magnetic permeability
increases monotonically inwards. The Bogomol'nyi solution (4.4) is obtained
from the general solution (4.5) and (4.6) by  letting $\alpha$ go to zero.

\beginsection 5. Massive Dilatons

It is widely believed by string theorists that the dilaton acquires a mass due
to non-perturbative effects connected with the breaking of supersymmetry.
It is therefore of interest to ask what the effect of a mass term for the
field $\phi$ might have on our solutions. In this section we shall simply
add in a mass term \lq\lq by hand \rq\rq . Thus we add to the Lagrangian in
(1.1)
a term $- {1 \over 2} m^2 \phi ^2$. One checks that the work above goes through
as long as one replaces $\nabla ^2$ by $\nabla ^2- m^2$ in (2.3) and (3.1).
However the solution (3.2) is no longer valid and we do not know how
to solve the new version of (3.1) exactly. The circularly symmetric solutions
may be treated  by regarding the radial coordinate
$r$ as a fictitious time variable. The equation for $\phi$ becomes
that of a particle  subject to  a time dependent frictional force and
moving in a potential $U (\phi)$

$$
- {1 \over 2} m^2 \phi ^2  + { 1 \over 2}  H^2 \exp ( 2 \tilde \kappa \phi )
\ . \eqno(5.1)
$$

Regularity at the origin implies that the radial derivative of $\phi$
vanishes there. A solution exists for each  value  of
$\phi$ at the origin.
If $m=0$ one has those solutions given in section 3 in which $\phi$ decreases
monotonically to minus infinity at infinity as
$$
\tilde \kappa \phi \sim -2 \log r. \eqno (5.2)
$$

For non-vanishing mass $m$ the behaviour depends on the ratio $m^2/(2e \tilde
\kappa ^2 H ^2) $  where $e$ is the base of natural logarithms.  If $ 0< m^2 /
(2 e \tilde \kappa ^2 H^2) < 1$ the potential $U(\phi)$ is a monotonic
increasing function of $\phi$ and for all values of $\phi(0)$,  $\phi$
decreases monotonically with $r$ and tends to  minus infinity  at infinity
$$
\phi \sim  - c \exp (mr) \eqno(5.3)
$$
where $c$ is a positive constant.

However if $m^2 /(2 e \tilde \kappa ^2H^2)$ is greater than unity then the
potential $U(\phi)$ has a local minimum and  maximum.
The behaviour of the solutions depends upon $\phi(0)$. If $\tilde \kappa
\phi(0)$ is positive and sufficiently large then $\phi$
decreases monotonically to minus infinity as before. However if $\tilde
\kappa \phi(0) $ lies in a finite interval bounded below by the smaller
solution $x$ of the equation:
$$
x=  {{\tilde \kappa ^2 H^2 } \over { m^2}} \exp 2x \eqno(5.4)
$$
then the solutions oscillate about the minimum with an  amplitude which
decreases to zero as
$r$ tends to infinity.
Finally if $\tilde \kappa \phi(0) <x$  the solutions decrease monotonically to
minus infinity.

Thus for given magnetic field $H$ there always exist solutions with finite
total flux. If however
the mass is large enough, one has  solutions in which $\phi$ tends a minimum
value dependent upon $H$ of the potential $U(\phi)$.

We have repeated these calculations for the monopole solutions of section 4
in the case of a massive dilaton. We find that  the qualitative
behaviour of the dilaton is very similar to that in the massless case.

One might imagine a cosmological scenario in which the dilaton is initially
massless at some high temperature and acquires
a  mass during a cosmological phase transition at a lower temperature.
If cosmic strings of the type we have described confining a finite flux
were initially present and the mass were large enough it seems from our
calculation that provided $\phi(0)$ took suitable values  the flux would become
unconfined. If this were true it might have important consequences for
magnetic monopoles. If flux was confined by strings at early times then
one might expect  magnetic monopoles, of the sort described in section 4, to
be found at the ends of flux tubes. These flux tubes should pull the monopoles
together and  cause their rapid  annihilation. At late times magnetic fields
would become unconfined. In this way one might have a natural solution to the
monopole problem. Clearly more work needs to be done to establish whether this
picture is really viable.

It is interesting to note that  dilaton electrodynamics
with an effective mass term has already been invoked [11] to account for a
possible
primordial magnetic field. It would interesting to investigate the relation
between that work and the monopoles and vortices described in this paper.

\noindent {\it Acknowledgement} We would like to thank Paul Shellard for some
helpful discussions and suggestions.

\vskip 1cm
\beginsection References

\medskip \item {[1]} W B Bonnor {\sl Proc Phys Soc Lond } {\bf A67} 225 (1954)
\medskip \item {[2]} M A Melvin {\sl Phys Lett } {\bf 8}  65 (1964)
\medskip \item {[3]} G W Gibbons \& K Maeda {\sl Nucl Phys } {\bf B298} 741-775
 (1988)
\medskip \item {[4]} F Dowker, J P Gauntlett, D A Kastor \& J Traschen hep-th
9309075
\medskip \item {[5]} T Maki \& K Shiraishi Akita Junior College preprint AJC
HEP-16 (1993)
\medskip \item {[6]}  G W Gibbons \& C G Wells DAMTP prepint R93/25 gr-qc
9310002
\medskip  \item {[7]}  G Lavrelashvili \& D Maison {\sl Phys Lett} {\bf B 295}
67 (1992)
\medskip \item {[8]} P Bizon {\sl Phys Rev} {\bf D47} 1656-1663 (1993)
\medskip \item {[9]} R Bartnik \& J McKinnon {\sl Phys Rev Lett} {\bf 61} 41
(1988)
\medskip \item {[10]} G W Gibbons in {\sl The Physical Universe} J D Barrow,
A B Henriques, M T V T Lago \& M S Longair {\sl Lecture Notes in Physics}
{\bf 383} Springer-Verlag (1991)
\medskip \item {[11]} B Ratra {\sl Ap J} {\bf 391} L1-L4 (1992)

\bye